\documentstyle[prl,aps,multicol,epsf]{revtex}
\begin{document}
\draft 
\title{Absence of Two-Dimensional Bragg Glasses} 
\author{Chen Zeng$^{a}$, P.L. Leath$^{a}$, and Daniel S. Fisher$^{b}$}
\address{(a) Department of Physics and Astronomy, Rutgers University, 
Piscataway, NJ 08854, USA} 
\address{(b) Department of Physics, Harvard University, 
Cambridge, MA 02138, USA}
\maketitle  

\widetext
\begin{abstract} 
The stability to dislocations of the elastic phase, or ``Bragg glass'', 
of a randomly pinned elastic medium in two dimensions is studied using
the minimum-cost-flow algorithm for a disordered fully-packed loop model.
The elastic phase is found to be unstable to dislocations due to the  quenched
disorder.  The energetics of dislocations are discussed within the framework of
renormalization group predictions as well as in terms of a domain wall picture.
\end{abstract} 

%\date\today

\begin{multicols}{2}
\narrowtext

%%%%%%%%%%%%%%%%%%%%%%%%%%%%%%%%%%%%%%%%%%%%%%%%%%%%%%%%%%%%%%%%%%%%%%%%%%%%%%
% Part (I).  What's a Bragg Glass? 
%%%%%%%%%%%%%%%%%%%%%%%%%%%%%%%%%%%%%%%%%%%%%%%%%%%%%%%%%%%%%%%%%%%%%%%%%%%%%% 

Randomly pinned elastic media are used to model various condensed-matter
systems with quenched disorder, including flux-line arrays in dirty type-II 
superconductors\cite{Blatter} and charge density waves\cite{Gruner}.
Although it is known\cite{Larkin} that these systems cannot exhibit
long-range translational order in less than four dimensions, the intriguing
possibility of a ``topologically ordered''  low-temperature phase remains
an open question\cite{BG_3d,4b,BG_23d}. It has been conjectured that such a phase,
with all dislocation loops bound, exists in three dimensions. Such 
a phase would be elastic and have power-law Bragg-like singularities
in its structure factor; it is often referred to as a ``Bragg'' or 
``elastic'' glass\cite{BG_3d,BG_23d}.  

%%%%%%%%%%%%%%%%%%%%%%%%%%%%%%%%%%%%%%%%%%%%%%%%%%%%%%%%%%%%%%%%%%%%%%%%%%%%%%
% Part (I). 2D Bragg Glass?  
%%%%%%%%%%%%%%%%%%%%%%%%%%%%%%%%%%%%%%%%%%%%%%%%%%%%%%%%%%%%%%%%%%%%%%%%%%%%%%

Whether or not unbound topological defects exist at low temperatures involves
a subtle balance between elastic-energy cost and disorder-energy gain\cite{4b}.   
In this letter, we analyze this issue for two-dimensional randomly pinned elastic
media at zero temperature by considering a $2d$-lattice model, viz.,
a fully-packed loop (FPL) model\cite{FPL} with quenched disorder. Exact ground
states of the FPL model are computed with and without topological defects,
which we refer to as dislocations. The {\em polynomial} optimization 
algorithm\cite{OP} that we use, minimum-cost-flow, enables us to study large
systems.

We focus on the energetics of a single dislocation pair in systems of size
$L\times L$. Our main conclusion is that the disorder energy gain of the
optimally placed pair dominates over the elastic energy cost with the 
results being consistent with theoretical predictions of $O(-\ln^{3/2}L)$
and $O(+\ln L)$ respectively for the two quantities. Dislocations 
therefore become unbound and proliferate causing the destruction of the 
Bragg glass. 

%%%%%%%%%%%%%%%%%%%%%%%%%%%%%%%%%%%%%%%%%%%%%%%%%%%%%%%%%%%%%%%%%%%%%%%%%%%%%%
\paragraph*{Model and algorithm}
%%%%%%%%%%%%%%%%%%%%%%%%%%%%%%%%%%%%%%%%%%%%%%%%%%%%%%%%%%%%%%%%%%%%%%%%%%%%%%
The FPL model is defined on a honeycomb lattice on which all configurations 
of occupied bonds which form closed loops and cover every site exactly once
are allowed, as shown in Fig.1(a). This model can be mapped to a solid-on-solid
surface model. Define integer heights at the centers of hexagons then orient
all bonds of the resulting triangular lattice such that elementary triangles 
pointing upward are circled clockwise; assign $1$ to the difference of 
neighboring heights along the oriented bonds if a loop is crossed and $-2$
otherwise. This yields single-valued heights up to an overall constant.
It can be seen that the smallest ``step'' of the surface is three, so that 
the effective potential on the surface is periodic in heights modulo 3. 

\begin{figure}
  \begin{center}
    \leavevmode
    \epsfxsize=8cm
    \epsffile{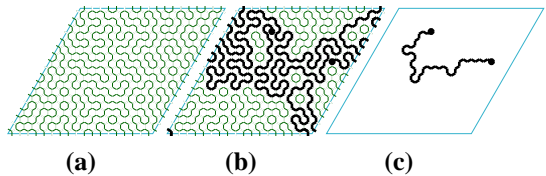}
  \end{center}
\caption{
The fully packed loop model with periodic boundary conditions imposed. 
The ground states with and without a pair of dislocations for one realization
of random bond weights are displayed in (b) and (a) respectively. 
The dislocations (solid dots) in (b) are connected by an open string 
(thick line) among the loops. The relevant physical object is, however,
the domain wall which is induced by the dislocations as shown in (c). This
domain wall represents the line of bond {\it differences} between the 
ground states (a) and (b).
}
\label{fig_1}
\end{figure}

Quenched disorder is introduced via random bond weights on the honeycomb lattice, 
chosen independently and uniformly from integers in the interval [-500, 500].
The energy is the sum of the bond weights along all loops and strings. The system
can be viewed as a surface in a $3d$ random medium that is periodic in the
height direction. Dislocations are added to the FPL model by ``violating'' 
the constraint. One dislocation pair is an open string in an otherwise
fully-packed system as shown in Fig.1(b). The height change along any
path encircling one end of the string is the Burgers charge $\pm 3$ of a
dislocation so that the heights  become multi-valued. Note that the 
configurations with and without a dislocation pair only differ on a
domain ``wall'',  as shown in Fig.1(c).

Finding ground states is equivalent to an integer minimum-cost flow problem
on a suitably designed graph; details can be found elsewhere\cite{OP,ZLF}. 

\paragraph*{Numerical results}  
For each disorder realization, the ground-state energies with and without
a dislocation pair and hence the defect energy, $E_d$, and also the connecting
wall (See Fig.1(c)) were computed. 

\begin{figure}
  \begin{center}
    \leavevmode
    \epsfxsize=8cm
    \epsffile{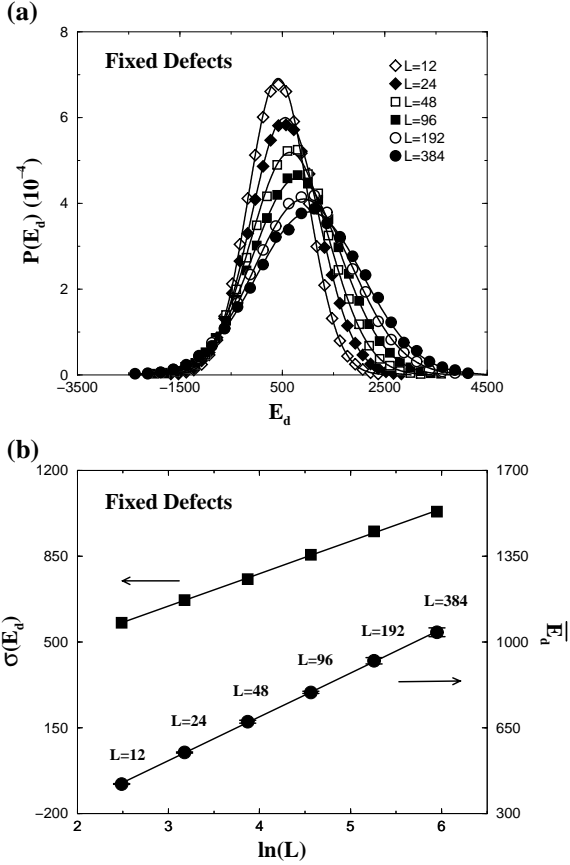}
  \end{center}
\caption{
Energetics of {\em fixed} defects: The probability distributions of the
energy of a {\em fixed} dislocation pair with separation $L/2$ for sample
sizes from $L=12$ to $L=384$ are shown in (a). The solid lines are gaussian
fits. The corresponding average defect energy $\overline{E_d}$ (solid circle)
and the root-mean-square (rms) width $\sigma(E_d)$ (solid square) are found
to scale with system size as $\ln L$ as shown in (b). The solid lines are
linear fits. 
}
\label{fig_2}
\end{figure}

We first held the dislocations {\it fixed} at two specific sites separated by
$L/2$ in $L\times L$ samples with $L=12,24,48,96,192$, and $384$, with at least
$10^4$ disorder realizations for each size. The probability distributions 
$P(E_d)$ of the defect energy are shown in Fig.2(a); they are well fit by 
gaussians for all sizes. The mean pair-energy is plotted versus $\ln L$ with 
a linear fit $\overline{E_d}=180(2)\ln L - 20(6)$. The root-mean-square (rms)
width $\sigma(E_d)$ is also shown in Fig.2(b), with a linear fit of
$\sigma(E_d) = 250(3) + 133(1) \ln L$. This implies a tail in $P(E_d)$ for
negative $E_d$ and suggests an energy {\em gain} from dislocations that 
can be optimized to take advantage of the negative part of the distribution. 

We have computed the {\em optimized} (lowest energy) dislocation pair energy
$E_d^{\rm min}$ for $L$ up to $480$ with $10^4-10^6$ samples for each size. 
The defect energy distribution $P(E_d^{\rm min})$ is no longer Gaussian,
indeed substantial asymmetry in $P(E_d^{\rm min})$ is seen in Fig.3(a). 
Moreover, in contrast to the case of fixed dislocations, 
$\overline{E_d^{\rm min}}$ is {\em negative} and {\em decreases} more
rapidly than $\ln L$ with increasing system size while the rms width 
$\sigma(E_d^{\rm min})$ increases less rapidly than $\ln L$. The linear fits 
shown in Fig.3(b) yield $[|\overline{E_d^{\rm min}}|]^{2/3}
= (43.80\pm0.31)\ln(L) + (24.03\pm0.14)$ and $[\sigma(E_d^{\rm min})]^2
= (21883\pm180)\ln(L) + (9013\pm802)$. 
This implies that
since almost all large systems have negative energy dislocation pairs, as is
evident in Fig.3(a) the FPL model is {\it unstable} against the
spontaneous appearance of dislocations. 

\begin{figure}
  \begin{center}
    \leavevmode
    \epsfxsize=8cm
    \epsffile{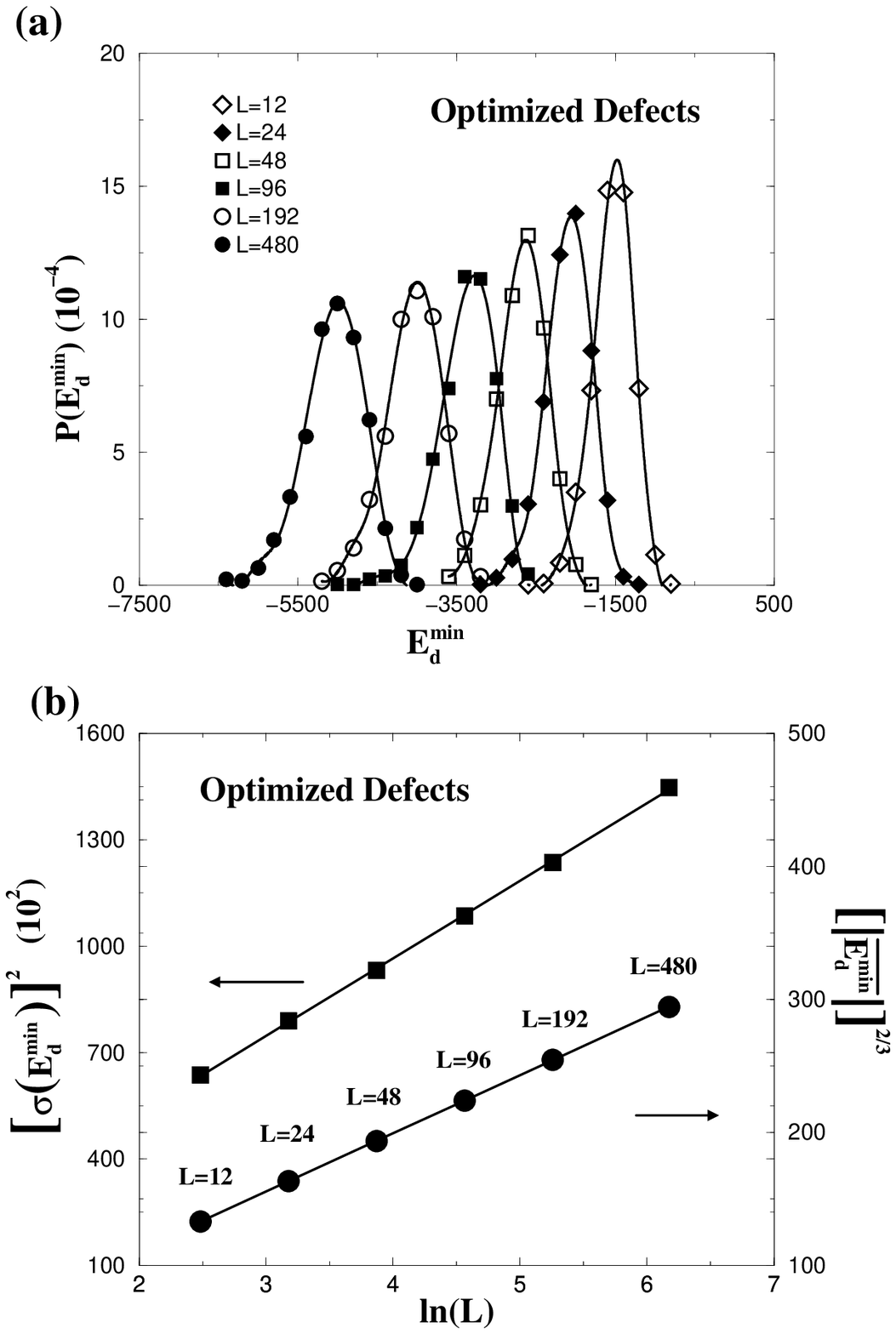}
  \end{center}
\caption{
Energetics of {\em optimized} defects: The defect energy probability
distributions for sample sizes from $L=12$ to $L=480$ are shown in (a).
The solid lines are guides to the eyes. Both the average defect energy
plotted as $[|\overline{E_d^{\rm min}}|]^{2/3}$ vs. $\ln L$
and the rms width plotted as $[\sigma(E_d^{\rm min})]^2$ vs. $\ln L$
are shown in (b). Solid lines are linear fits.
}
\label{fig_3}
\end{figure}

\paragraph*{Continuum models}
To shed some light on the observed defect energetics, and to make contact with
analytic work, we consider coarse-grained continuum approximations to the
random FPL model.  In the absence of dislocations, the surface has a stiffness
caused by the inability of a tilted and hence more highly constrained surface
to take as much advantage of the low weight bonds as a flatter surface. In
addition, the random bonds couple to ${\bf \nabla}h$ as well as to $h$ modulo
$b=3$. An appropriate effective
Hamiltonian is thus

\begin{equation}
H=\int d{\bf r}
\left[
\frac{K}{2} (\nabla h)^2 
- {\bf f}\cdot\nabla h  
- w\cos\left(\frac{2\pi}{3} h - \gamma\right) 
\right]
\label{eq1}
\end{equation}
with ${\bf f}\equiv{\bf f}({\bf r})$ locally random with variance $\Delta$,
and  $\gamma\equiv\gamma({\bf r})=0$, $\pm \frac{2\pi}{3}$ with equal probability.

A related and well studied model (CDW) \cite{RG} has, instead, the $\{\gamma({\bf
r})\}$ being uniformly distributed random variables in $[0,2\pi]$.  While it
is not clear a priori that these models are in the same universality class,
both the following, and an RG analysis, imply they are: shifting $h({\bf
r})$ by $\frac{3}{2\pi}\gamma({\bf r})$ changes the $\{f({\bf r})\}$ 
introducing short range correlations that are different in the two
models, but these should,  be irrelevant  on large scales of $h$ and $L$.
The CDW model has an elastic glass phase for $T<T_g$ below which $w$ is relevant
and renormalizes to a $T$-dependent fixed point while $\Delta$ grows as $\ln L$
yielding height variations
$\overline{<[h({\bf r})-h(0)]^2>}\approx \frac{b^2}{\pi}\Upsilon\ln^2
r$\cite{RG}.

An approximate functional RG analysis yields a similar structure at all
temperatures with a universal $T=0$ limit $\Upsilon_0$ of the coefficient
$\Upsilon(T)$ \cite{gaussian}. On large scales, the behavior is dominated by the
competition between the random stressing ${\bf f}({\bf r})$, and the stiffness
$K$, with the random force correlations effectively
$\overline{f_i({\bf q})f_j({\bf -q})} = - C\delta_{ij} \ln q^2$ for
small wave vectors $q$. We can thus work with the simple purely random-force
limit of Eq.~(\ref{eq1}) with $w=0$. We then immediately conclude that
$\Upsilon_0=\frac{C}{K^2 b^2}$. 
 
In the presence of dislocations $h({\bf r})$ becomes multi-valued. 
It can be decomposed into two parts $h = h_e + h_d$, with  $h_e$, the smooth 
elastic distortion  while the singular function $h_d$, has a cut
connecting the two dislocations at ${\bf r}_1$ and ${\bf r}_2$ with Burgers 
charge $b_1 = b = 3$ and $b_2 =-b=-3$ with $\nabla \times \nabla h_d =
\sum_{i=1,2} b_i  \delta^2({\bf r}-{\bf r}_i)$ and $\nabla^2h_d=0$. The energy of
a dislocation pair is
%!!!!!!!!!!!!!!!!!!!
\begin{eqnarray}
E_d
&=&
{K\over2} \int d^2{\bf r} |\nabla h_d|^2
-\int d^2{\bf r} ({\bf f}^T\cdot\nabla h_d)
\label{eq2} \\  
&\approx& 
{Kb^2\over2\pi} \ln |{\bf r}_1-{\bf r}_2|  
-b\left\lbrack g({\bf r}_1)-g({\bf r}_2)\right\rbrack  
\;\; . 
\label{eq3} 
\end{eqnarray} 
%!!!!!!!!!!!!!!!!!!! 
where the static random force field has been decomposed into longitudinal 
${\bf f}^L=\nabla u(\bf r)$ and  transverse ${\bf f}^T= \nabla\times g(\bf r)$
components. Since $u(\bf r)$, $g(\bf r)$, and $\nabla h_d$ are continuous 
across the cut while $h_d$ jumps by $b$, ${\bf f}^L$ makes no contribution 
to $E_d$ and Eq.(\ref{eq3}) follows by integration by parts. The first term is 
the standard elastic cost and the second the disorder gain so that $g({\bf r})$
is the {\it potential} felt by the dislocations. Its variance is $S_g \equiv
\overline{g({\bf q})g({\bf -q})} = - C  q^{-2}\ln q^2$ so that the elastic
surface without dislocations and the dislocation potential have the same 
statistics after a rescaling: $S_h\equiv\overline{h({\bf q})h({\bf -q})}
\approx S_g/K^2$. 

In terms of  the statistical properties of the dislocation potential our
numerical results can be understood. We first discuss fixed dislocation pairs.
It has been argued\cite{gaussian} from a statistical symmetry (of the CDW model) 
that on large scales $f({\bf r})$ and hence the potential $g({\bf r})$ and $E_d$ 
are gaussian. The shape of our computed defect energy distribution, its mean 
and its variance all agree with predictions from Eq.~(\ref{eq3}), $\overline{E_d}
\approx ({Kb^2}/{2\pi}) \ln L $ ---an exact result for the CDW model---and 
$\sigma(E_d) \approx \sqrt{b^2C/\pi}\ln L$ with $K=126(2)$ and $C=6174(93)$ 
yielding $C/K^2 = 0.389(11)$.  We also measured the variance of the height 
of the surface without dislocations finding a good fit to $\sigma^2(h) =
0.061(2)\ln^2 L + 0.477(7)\ln L + 0.765(13)$. The predicted coefficient of 
the $\ln^2 L$ is $C/2\pi K^2$ yielding $C/K^2 = 0.387(12)$. The agreement
between the two estimates of $C/K^2$ further supports the validity of the 
random  force model, including, in particular, the equality between the
longitudinal and transverse parts of ${\bf f(r)}$ 	

We now turn to optimized dislocations.  If the first term in Eq.~(\ref{eq3})
can be ignored, the energy of a dislocation pair will be lowest if the 
dislocations are at the minimum ($g_{min}$) and maximum 
($g_{max}$) of the random potential, $g({\bf r})$.
The distribution of the extrema of potentials like $g({\bf r})$ whose 
variance grows as a power of $\ln L$, can be semi-quantitatively understood 
by thinking of iteratively optimizing over each factor of two in length scale,
with the component of the random potential on scale $l$ being essentially the
contribution from Fourier components with $\frac{\sqrt{2}\pi}{l}<|q_x | ,|q_y 
|<\frac{2\sqrt{2}\pi}{l}$. If scale $l$ gives rise to a contribution to the 
variance of $g$ of order $(\ln l)^{2\alpha}$---with $\alpha=\frac{1}{2}$ in our 
case---, then a {\it typical} $g({\bf r})$ is the incoherent logarithmic sum 
over all scales, i.e. of order $(\ln l)^{\alpha +1/2}$. The maximum of $g$ can 
be found, heuristically, by maximizing over the four points at scale $l=1$ in 
each square of scale 2, then maximizing over the four scale 2 maxima in each 
scale 4 square, etc. Since the scale $l$ structure of $g$ is weakly correlated
over scales much longer than $l$, a crude approximation is to ignore these 
correlations whereby one stage of optimization adds of order $(\ln l)^\alpha$
to the local scale $l$ maximum\cite{extremal}. Thus scales should be summed over
{\it coherently} yielding
$g_{max}\sim(\ln L)^{\alpha +1}$.
From this hierarchical construction, it can be seen that
the variance of $g_{\rm max}$ is dominated by the largest scales,  so that
$\sigma(g_{\rm max})\sim (\ln L)^\alpha$. In our case, we thus expect 
$\overline{g_{\rm max}-g_{\rm min}}\sim(\ln L)^{3/2}$
which indeed dominates over the $\ln L$
elastic energy term in Eq.~(\ref{eq3}). Hence we expect $\overline{E_d^{\rm min}}
\approx - Ab\sqrt{C}(\ln L)^{3/2}$ and 
$\sigma(E_d^{\rm min}) \approx Bb\sqrt{C}(\ln L)^{1/2}$ 
with some coefficients $A$ and $B$. 

The linear fits in Fig.3(b) give 
$A\approx 1.23(1)$ and $B\approx0.56(1)$ using the $C$ computed earlier. 
If the elastic part $\frac{Kb^2}{2\pi}\ln L$ is 
subtracted from $\overline{E_d^{\rm min}}$ by fitting the difference
$\overline{E_d^{\rm min}(L)} - \overline{E_d^{\rm min}(L/2)}
\ {\rm to }\ -(3\ln 2/2)Ab\sqrt{C}(\ln L)^{1/2} + const.$, this yields  a very
comparable
$A\approx 1.17(5)$. Extremal heights $\overline{h_e^{\rm min}}\equiv 
\overline{h_{min}-h_{max}}$ of the elastic surface without dislocations 
and optimal dislocation energies can also be used to extract $K$ 
via $\overline{E_d^{\rm min}} - bK \overline{h_e^{\rm min}} 
\approx \frac{Kb^2}{2\pi}\ln L + const.$. This yields $K\approx 114(1)$ in not
unreasonable  agreement with the $K\approx 126(2)$ from the {\it fixed}
dislocation pairs.

An upper bound on $\overline{g_{\rm max}-g_{\rm min}}$ can be simply obtained by noting 
that ${\rm Prob }[g_{\rm max}>M]\leq \sum_{{\bf r}}{\rm Prob}[g({\bf r})>M]$ 
%!!!!!!!!!!!!!!!!!!!! 
%\begin{equation}
%{\rm Prob }[g_{\rm max}>M]\leq \sum_{{\bf r}}{\rm Prob}[g({\bf r})>M]
%\label{eq4}
%\end{equation}
%!!!!!!!!!!!!!!!!!!!!
so that, with $L^2$ points, the median $g_{\rm max}$ is less than the $M$ for
which the right hand side, $L^2 {\rm Prob}[g({\bf r})>M]$ for fixed ${\bf
r}$, equals $\frac{1}{2}$.  If $g({\bf r})$ is gaussian sufficiently far into
the tail of its distribution, this yields $\overline{g_{\rm max}-g_{\rm min}}\leq
\sqrt{C}\sqrt{\frac{8}{\pi}}(\ln L)^{\frac{3}{2}}$ so that $A\leq\sqrt{\frac{8}{\pi}}
=1.5957...$. The hierarchical optimization described above
suggests that $A$  might saturate this bound. To test the universality of $A$ 
and this issue, we have measured the distributions 
of the extrema of the heights of the FPL surface without dislocations 
and several simulated exactly gaussian random surfaces with different $\alpha$.
Good fits are found to $(\ln L/a_\alpha)^{1+\alpha}$ with $a_\alpha$ a cutoff,
yielding $A_{\rm FPL\ heights}\approx 1.45(3)$, and 
$A_{\rm gaussian}\approx 1.517(3), 1.307(6),  1.168(6)$, and
$1.064(5)$ for 
$\alpha=0,1/2,1$, and $3/2$ respectively.  Similarly
from the variances we obtain
$B_{\rm FPL\ heights}\approx 0.67(2)$, and 
$B_{\rm gaussian}\approx 0.475(4), 0.637(7), 0.730(9)$, 
and $0.814(4)$ for $\alpha=0,1/2,1$, and $3/2$ respectively. 

We first note that all the extracted $A$'s satisfy $A\leq\sqrt{8/\pi}$, and
there appears to be a systematic trend for $A(\alpha)$ to decrease as $\alpha$
increases; if this is true then it is most likely that $A$ is strictly less
than the bound for all $\alpha>-\frac{1}{2}$ (for $\alpha<-\frac{1}{2}$
gaussian surfaces the variance saturates for large $L$ and the extrema grow as
$\sqrt{\ln L}$). The values of $A$ for the nominally $\alpha=\frac{1}{2}$
cases, $1.17$-$1.23$, $1.45$ and $1.307$ for the optimal dislocations, extrema heights,
and the gaussian surface respectively, differ by substantially more than the
apparent statistical errors as do the $B$'s, $0.56$, $0.67$, $0.637$ respectively. 
But given the narrow range available of $\ln L$ in spite of a large range of
sizes and, as importantly, the lack of understanding of corrections to
scaling, these results are certainly consistent with universal values of $A$
and $B$ for $\alpha=\frac{1}{2}$.  At this point, however, understanding
whether this is in fact the case, and also whether $A$ for gaussian surfaces
depends on $\alpha$, must wait for better theoretical understanding.

%%%%%%%%%%%%%%%%%%%%%%%%%%%%%%%%%%%%%%%%%%%%%%%%%%%%%%%%%%%%%%%%%%%%%%%%%%%%%
% Conclusions. 
%%%%%%%%%%%%%%%%%%%%%%%%%%%%%%%%%%%%%%%%%%%%%%%%%%%%%%%%%%%%%%%%%%%%%%%%%%%%%

Overall, we have found rather good agreement for a variety of large scale
quantities with the RG prediction of equivalence at long scales of the FPL model
and a random force model. Although extracting reliable exponents of $\ln L$
is not possible (especially with logarithmic corrections to scaling) the
fact that the {\it coefficients} and ratios between these---extracted several
ways---are in reasonable agreement is a more stringent test. But even if the
random force equivalence is {\it not} valid, the data of Fig.3 clearly indicate
the instability of large systems to   dislocation pairs. With no
restrictions on their number, dislocations will proliferate thereby
driving the elastic constant $K$ to zero. 

We conclude with an alternate way to understand the structure of excitations
in the elastic glass, via a picture developed for the three-dimensional
case\cite{4b}. The basic excitations from a ground state are fractal domain walls
surrounding regions in which $h$ changes by $b$.  Their fractal dimension,
$d_w$,  will be the same as that for the forced open wall that connects a pair
of dislocations (Fig.1(c)) for which we find  $d_w=1.28(3)$ for fixed pairs and
1.30(3) for optimized pairs.  [These contrast strongly with the connecting
{\it strings} in the loop model which have $d_s=1.75(3)$ and 1.74(3)
respectively, very close to the value in the non-random loop model\cite{FPL}]. 
The energy of a scale $L$ wall constrained only on scale $L$ is
predicted to vary by of order
$\sqrt{\ln L}$ but have mean independent of $L$. The incoherent logarithmic
addition  over all scales then yields variations of the fixed-end open
domain wall energy, of order $\ln L$ and a mean of the same order---as
found.  But if the end positions can adjust to lower the wall energy near
the dislocation at each scale, the energies add up {\it coherently} resulting in
the $- \ln^{3/2} L $ mean optimal dislocation pair energy with order
$\sqrt{\ln L}$ around the mean variations being dominated by the largest scale,
in an analogous way to the extrema of the random potential $g({\bf r})$ of the
random force picture.  Since the defect energy in the domain wall picture is
concentrated on the wall, while it is spread out over a region of area $O(L^2)$
in the random force model, it is surprising that these yield the same
predictions! But the fact that our results agree well with the domain wall
picture in $2D$ lends strong support to the validity of the analogous picture in
the $3D$ case for which  it has been used to conclude that the $3D$
elastic glass phase is stable to dislocation loops\cite{4b}.

We thank J. Kondev and C.L. Henley for useful discussions. 
This work has been supported in part by the National Science Foundation via
grants DMR 9630064, DMS 9304586 and Harvard University MRSEC.

\end{multicols}
\end{document}